\documentclass[aps,prl,reprint,superscriptaddress, showpacs]{revtex4-1}

\usepackage{graphicx}
\usepackage{dcolumn}
\usepackage{bm}
\usepackage{times}
\usepackage{color}
\usepackage{braket}

\begin{document}

\bibliographystyle{unsrt}
\preprint{C. Kloeffel {\em et al.} \today}

\title{Controlling the Interaction of Electron and Nuclear Spins in a Tunnel-Coupled Quantum Dot}

\author{C. Kloeffel}
\affiliation{Department of Physics, University of Basel, Klingelbergstrasse 82, CH-4056 Basel, Switzerland}

\author{P. A. Dalgarno}
\affiliation{School of Engineering and Physical Sciences, Heriot-Watt University, Edinburgh EH14 4AS, UK}

\author{B. Urbaszek}
\affiliation{Universit\'{e} de Toulouse, INSA-CNRS-UPS, LPCNO, 135 Avenue de Rangueil, 31077 Toulouse, France}

\author{B. D. Gerardot}
\affiliation{School of Engineering and Physical Sciences, Heriot-Watt University, Edinburgh EH14 4AS, UK}

\author{D. Brunner}
\affiliation{School of Engineering and Physical Sciences, Heriot-Watt University, Edinburgh EH14 4AS, UK}

\author{P. M. Petroff}
\affiliation{Materials Department, University of California, Santa Barbara, California 93106, USA}

\author{D. Loss}
\affiliation{Department of Physics, University of Basel, Klingelbergstrasse 82, CH-4056 Basel, Switzerland}

\author{R. J. Warburton}
\affiliation{Department of Physics, University of Basel, Klingelbergstrasse 82, CH-4056 Basel, Switzerland}
\affiliation{School of Engineering and Physical Sciences, Heriot-Watt University, Edinburgh EH14 4AS, UK}

\date{\today}

\pacs{73.21.La, 78.67.Hc}


\begin{abstract}
We present a technique for manipulating the nuclear spins and the emission polarization from a single \mbox{optically} active quantum dot. When the quantum dot is tunnel coupled to a Fermi sea, we have discovered a natural cycle in which an electron spin is repeatedly created with resonant optical excitation. The spontaneous emission polarization and the nuclear spin polarization exhibit a bistability. For a $\sigma^{+}$ pump, the emission switches from $\sigma^{+}$ to $\sigma^{-}$ at a particular detuning of the laser. Simultaneously, the nuclear spin polarization switches from positive to negative. Away from the bistability, the nuclear spin polarization can be changed continuously from negative to positive, allowing precise control via the laser wavelength.
\end{abstract}

\maketitle

Semiconductor quantum dots are very attractive for applications as qubits \cite{losspra} and sources of quantum light \cite{pmichlerscience, akopianprl, salternature}. Versatile materials are the III-V semiconductors, notably GaAs which has established itself as the workhorse material. A significant property is that all the Ga, As, and In isotopes have large nuclear spins. In a typical quantum dot there is an intermediate number of atoms, too large to use each nuclear spin as a resource yet too small for efficient cancellation in the total spin, and noise in the nuclear spins limits the electron spin coherence to just $\sim 10$ ns through the hyperfine interaction \cite{khaetskiiprl, mhmikkelsennatphys, xxunaturephys}. An emerging theme is that the nuclear spin noise may be reduced by narrowing the distribution \cite{coishprb, agreilichscience, xxunature2009, Bluhm} and that the nuclear spin ensemble may represent as much opportunity as trouble. Currently, schemes exist to tune both the optical transition energy \cite{lattanaturephysics} and the selection rules \cite{tbelhadjphysrevlett} of a quantum dot in situ, but presently, the possibilities of using nuclear spins beneficially are limited.

We present here a new control over the electron-nuclear spin interaction on driving an optical transition resonantly. Dynamic nuclear polarization at the single quantum dot level is established \cite{aitartakovskiiprl, pmaletinskyphysrevlett, mnmakhoninapl, eachekhovichprl, fklotzphysrevb}. The crucial advance here is to operate in the tunneling regime \cite{eachekhovichprl, fklotzphysrevb} where we discover a natural cycle. There are two interrelated features. First, spontaneous emission following resonant excitation either preserves the circular polarization of the source or inverts it. For instance, with a $\sigma^{+}$ pump, we can switch from predominantly $\sigma^{+}$ to $\sigma^{-}$ emission either with a small change in pump wavelength or device bias allowing the polarization of a single photon source to be controlled in situ. Secondly, the resonant excitation creates a large nuclear spin polarization which changes sign abruptly at the bistability, a new feature compared to the bistabilities following nonresonant optical excitation \cite{aitartakovskiiprl, braunprb}. At smaller laser wavelengths, the nuclear spin polarization changes monotonically from a large negative value to a large positive value. This bidirectional tuning is demonstrated here at low magnetic fields (0.5 T), and complements the optical dragging effect at high magnetic fields \cite{lattanaturephysics}. Control of the nuclear spins via the optical wavelength is a powerful route to narrowing the distribution \cite{lattanaturephysics} and to tuning the quantum dot exciton over tens of $\mu$eV.

Our experiments use a field effect device in which InGaAs self-assembled quantum dots are in tunnel contact with an $n^{+}$ GaAs Fermi sea via a 25 nm thick GaAs tunnel barrier \cite{padalgarnophysrevlett}. A voltage is applied to a Schottky contact on the sample surface, 150 nm above the quantum dot layer, at 4.2 K. Photoluminescence (PL) is excited either nonresonantly at 830 nm wavelength, or resonantly using 13 kW/cm$^{2}$ from a tunable narrowband cw laser. The PL is dispersed with a monochromator and detected with a CCD array detector, a system with resolution 50 $\mu$eV. The polarization of excitation and collection are independently controlled. A small magnetic field, $B_z = +0.5\mbox{ T}$, is applied along the growth $\vec{z}$ direction.

\begin{figure}[t]
\vspace{0.0cm}                                 
\centering                                     
\includegraphics[width=0.955\linewidth]{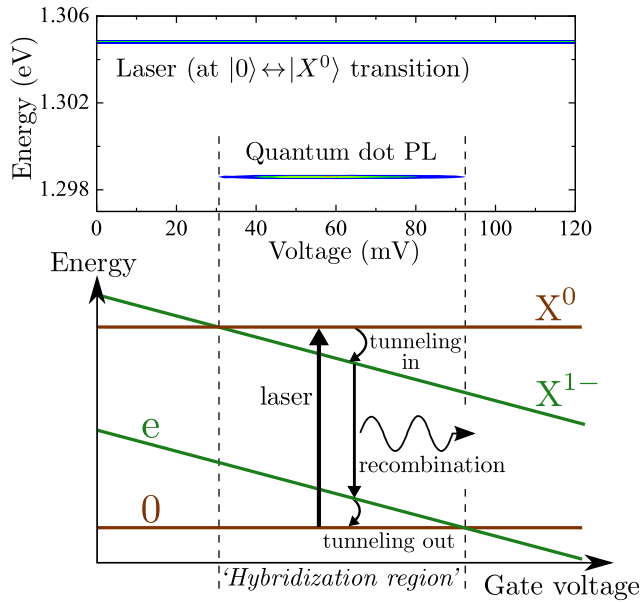}   
\vspace{0.0cm} 
\caption[]{Top: Photoluminescence (PL) at 4.2 K from a single quantum dot versus bias driven with excitation at the $X^{0}$ energy. $X^{1-}$ PL appears in a narrow range of voltage, the hybridization regime. Bottom: Energy dependence versus bias for the quantum dot vacuum state $\ket{0}$ and the single electron state $\ket{e}$, showing a crossing where the ground state changes. $X^{0}$ and $X^{1-}$ cross at lower bias on account of the hole. Within the hybridization region, automatic cycling takes place when a laser is tuned to the $\ket{0} \leftrightarrow \ket{X^{0}}$ transition. An electron tunneling from the Fermi sea turns the $\ket{X^{0}}$ into $\ket{X^{1-}}$; recombination leaves the system in state $\ket{e}$; tunneling out returns the dot to $\ket{0}$.}  
\label{figure1}                                       
\end{figure}

Excited nonresonantly, the photoluminescence from a single quantum dot shows a clear charging step from the neutral exciton, $X^{0}$, to the negatively charged trion, $X^{1-}$. The energies of the initial states $\ket{X^{0}}$ and $\ket{X^{1-}}$, and their corresponding final states, $\ket{0}$ (vacuum) and $\ket{e}$ (single electron), as a function of gate voltage are shown in Fig.\ \ref{figure1}. In the final states (no hole present), the ground state charges from $\ket{0}$ to $\ket{e}$ at a more positive voltage than the change in the initial states from $\ket{X^{0}}$ to $\ket{X^{1-}}$ \cite{padalgarnophysrevlett}, a consequence of the different electron-hole and electron-electron Coulomb energies. A ``hybridization region" is created, a voltage region in which both excitons are tunnel coupled to the Fermi sea, $X^{0}$ in the initial state, $X^{1-}$ in the final state \cite{padalgarnophysrevlett}. We show here that this region is ideal for controlling the electron-nuclear spin interaction. 

Figure\ \ref{figure1} shows the result of pumping the $\ket{0}\leftrightarrow\ket{X^{0}}$ transition of a single quantum dot. Over a small region of voltage, $X^{1-}$ PL is observed, redshifted by 6 meV with respect to the laser. A comparison with the nonresonantly excited PL demonstrates that this region corresponds to the low bias edge of the $X^{1-}$ plateau, i.e.\ the hybridization region, and that the resonantly excited PL has the $X^{0}$ energy. In terms of the level diagram in Fig.\ \ref{figure1}, the dot is initially in the vacuum state $\ket{0}$. The laser then creates an $X^{0}$, which, although neutral, is unstable with respect to tunneling. Electron tunneling {\em into} the dot (time scale $\sim 50$ ps, considerably shorter than the radiative lifetime of $\sim 1$ ns) creates an $X^{1-}$ which then recombines. After spontaneous emission, the dot is in the $\ket{e}$ state. Now that the hole has disappeared, this state is also unstable with respect to tunneling: electron tunneling {\em out} of the dot (time scale $\sim 10$ ps) returns the dot to $\ket{0}$ whereupon the process can be repeated. This cycle offers a number of attractive features. First, the $X^{0}$ spin is determined by the polarization of the laser through the optical selection rules. Second, the cycle round-trip time is small, just $\sim 1$ ns, limited only by spontaneous emission. Third, the redshift of the PL with respect to the excitation makes it easy to distinguish spontaneous emission from scattered laser light even though one of the transitions is driven resonantly. The PL is useful in its own right as an antibunched source. It also provides an in situ monitor of the nuclear spin polarization through the Overhauser shift. Finally, the process can be described quantitatively with no ad hoc assumptions.

\begin{figure}[t]
\vspace{0.0cm}                                 
\centering                                     
\includegraphics[width=1.00\linewidth]{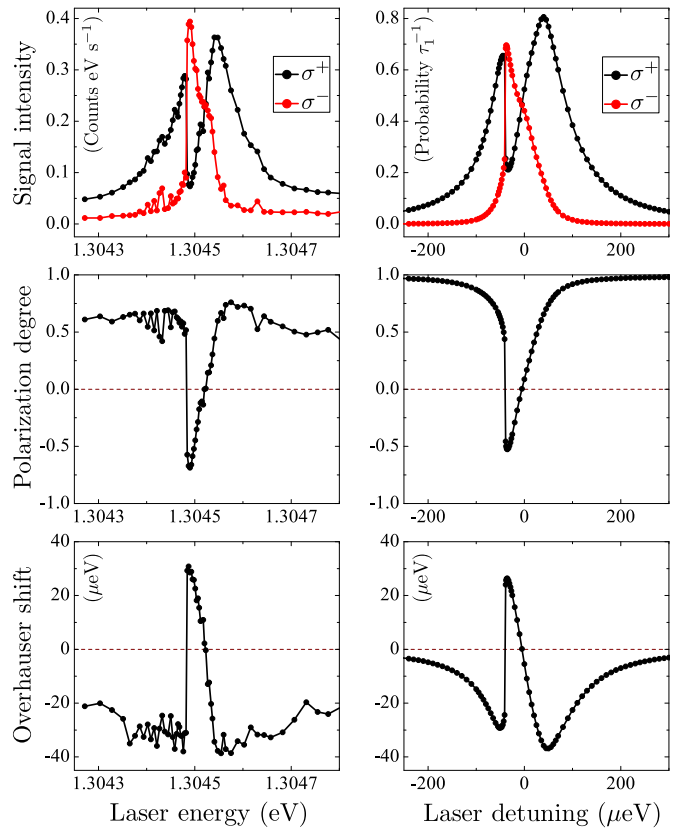}   
\vspace{0.0cm} 
\caption[]{Left (right) panels: Experimentally measured (calculated) signal intensity, polarization degree, and Overhauser shift versus laser energy (laser detuning) for a $\sigma^{+}$ pump and an external field of $+0.5\mbox{ T}$ at fixed bias in the center of the hybridization region. In the experiment, the laser is tuned close to the $\ket{0} \leftrightarrow \ket{X^{0}}$ transition and the dot (at 4.2 K) is the same as in Fig.\ \ref{figure1}.}  
\label{figure2}                                       
\end{figure}

The main experiment consists of monitoring the $X^{1-}$ PL as a function of laser detuning with respect to the $X^{0}$ transition for a constant pump polarization, e.g.\ $\sigma^{+}$. A PL spectrum is recorded for both $\sigma^{+}$ and $\sigma^{-}$ polarizations. These counts-energy spectra are fitted to Lorentzians \cite{epaps}, yielding both the signal energy $E(\sigma^{+/-})$ (center of Lorentzian), and the signal intensity $S(\sigma^{+/-})$ (area under Lorentzian). Figure \ref{figure2} shows both $S(\sigma^{+})$ and $S(\sigma^{-})$ for a $\sigma^{+}$ pump, and the associated polarization degree
\begin{displaymath}
P = \frac{S(\sigma^+)-S(\sigma^-)}{S(\sigma^+)+S(\sigma^-)}.
\end{displaymath} 
At large negative and positive detunings, the PL has largely $\sigma^{+}$ character with $P$ up to $0.76 \pm 0.05$. This is the intuitive result from the selection rules. Absorption of a $\sigma^{+}$ photon with spin angular momentum $+\hbar$ along $\vec{z}$ creates an $\ket{\Uparrow \downarrow}$ exciton consisting of a heavy hole $\Uparrow$ with angular momentum $\vec{z}$ projection $+\frac{3}{2}\hbar$ and an electron $\downarrow$ with $-\frac{1}{2}\hbar$. An electron tunnels in to form the $X^{1-}$ exciton, $\ket{\Uparrow \downarrow \uparrow}$. Hole spin relaxation is slow compared to recombination \cite{dheissphysrevb, bdgerardotnature} such that recombination $\ket{\Uparrow \downarrow \uparrow} \rightarrow \ket{\uparrow}$ creates a $\sigma^{+}$ photon. The counterintuitive result in Fig.\ \ref{figure2} is that close to the center of the resonance, the PL has an {\em inverted} polarization degree, with $P\sim -0.7$. Strikingly, $P$ changes abruptly at a particular detuning.

An indicator that the nuclear spins are involved is provided by the Overhauser shift 
\begin{displaymath}
\Delta_{\rm n} = E(\sigma^+) - E(\sigma^-) - g_{\rm X} \mu_{\rm B} B_z.
\end{displaymath} 
$\Delta_{\rm n}$ is interpreted as an energy shift of the unpaired electron spin in the $X^{1-}$ final state arising from the nuclear spin polarization along $\vec{z}$. Its determination requires a knowledge of the exciton g-factor, and we measure $g_X=1.55 \pm 0.10$ as described in \cite{epaps}. Close to the center of the resonance we now find that $\Delta_{\rm n}$ switches sign exactly at the point where $P$ switches sign. The Overhauser shift is related to the average nuclear spin $\vec{z}$ projection $\langle I_z \rangle$ (in units of $\hbar$) through $\Delta_{\rm n} \simeq - A \langle I_z \rangle$ \cite{epaps}. Taking the coupling constant $A \approx 90\mbox{ $\mu$eV}$, an averaged value for In$_{\rm 0.5}$Ga$_{\rm 0.5}$As \cite{wacphysstatussolidib}, we find that $\langle I_z \rangle \approx +0.36 \leftrightarrow -0.36$. Full polarization corresponds to $\langle I_z \rangle = \pm 2.25$, where $I=2.25$ is the average nuclear spin quantum number in the dot.    

\begin{figure}[t]
\vspace{0.0cm}                                 
\centering                                     
\includegraphics[width=0.905\linewidth]{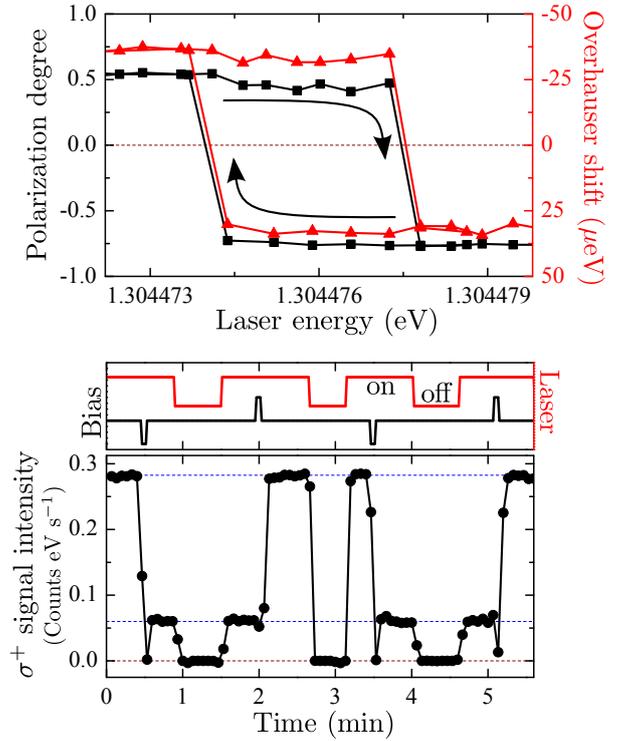}   
\vspace{0.0cm} 
\caption[]{ Top: Bistability between state I ($\sigma^{+}$ PL: nuclear spins up) and state II ($\sigma^{-}$ PL: nuclear spins down). $P$ and $\Delta_{\rm n}$ were measured as the laser was tuned. The laser was blocked for 30 s between each point. Bottom: Demonstration of switching with gate voltage pulses (40 mV, 5 s duration) by measuring the $\sigma^{+}$ PL; strong PL signifying state I, weak PL signifying state II. In between voltage pulses, the laser was turned off. Both curves were recorded at $+0.5\mbox{ T}$, 4.2 K, using a $\sigma^{+}$ pump and the same dot as in Figs.\ \ref{figure1} and \ref{figure2}.}  
\label{figure3}                                       
\end{figure}

The abrupt jump in $P$ corresponds to a bistability. With $\sigma^{+}$ excitation, in state I (II) the dot emits $\sigma^{+}$ ($\sigma^{-}$) photons and the nuclear spins point up (down). The bistability is demonstrated clearly in the hysteresis curve of Fig.\ \ref{figure3} (top). In this case, the laser energy was tuned in fine steps (less than 0.5 $\mu$eV), blocking the laser path for about 30 s between each data point during which time the state of the system was always preserved. At more positive laser detunings, the polarization degree $P$ and the nuclear spin polarization are continuous monotonic functions of detuning, changing from large negative to large positive values. Correspondingly, $\Delta_{\rm n}$ goes smoothly from $+30$ to $-35$ $\mu$eV. The total electron Zeeman splitting, $g_{\rm e}^{\rm eff}\mu_{\rm B} B_z = g_{\rm e} \mu_{\rm B} B_z + A\langle I_z \rangle$, changes  sign at the bistability, followed by continuous tuning from $-45$ to $+20\mbox{ $\mu$eV}$ (tuning of effective electron g-factor $g_{\rm e}^{\rm eff}$ from $-1.6$ to $+0.7$). 

To switch from state I to state II, it is more convenient to change the gate voltage than the laser wavelength. We have achieved this by exploiting the Stark effect of the exciton energy. Figure \ref{figure3} (bottom) demonstrates controlled switching between state I and II by applying voltage pulses to the gate, monitoring the state of the system via the $\sigma^{+}$ PL. The system is initially in state I. It is forced into state II with a negative voltage pulse, equivalent to moving the laser energy up and back down again. This results in a lower $\sigma^{+}$ PL, the signature of state II. Analogously, we can switch the system back into state I with a positive voltage pulse. In between these voltage pulses, the laser is turned off. When it is turned back on again $\sim 30\mbox{ s}$ later, the system always adopts its original state, demonstrating a slow I-II relaxation rate ($<0.1 \mbox{ s}^{-1}$).

We present a quantitative model to describe these results. The two crucial ingredients are first, a coherent coupling between $\ket{\Uparrow \downarrow}$ and $\ket{\Downarrow \uparrow}$, the so-called fine structure which arises from the anisotropic part of the electron-hole exchange, and second, a hyperfine coupling between the nuclear spins and the unpaired electron spin. A full description of the model is given in \cite{epaps}.

\begin{figure}[t]
\vspace{0.0cm}                                 
\centering                                     
\includegraphics[width=0.93\linewidth]{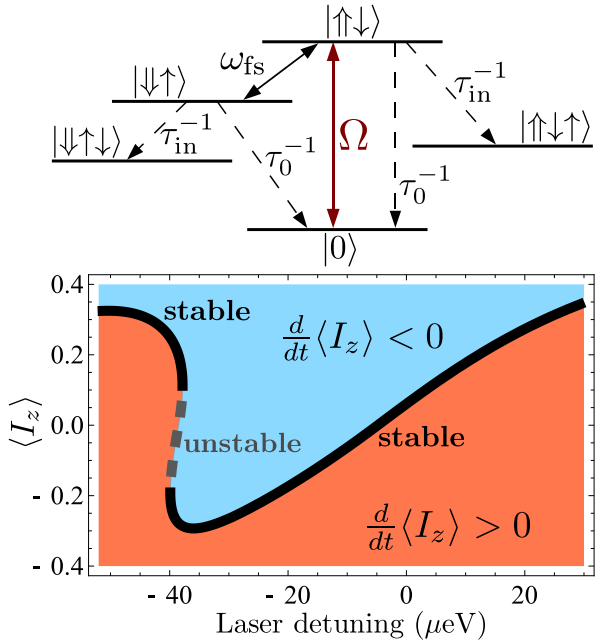}   
\vspace{0.0cm} 
\caption[]{Top: The five quantum states in the simulation showing an optical coupling (Rabi energy $\hbar \Omega$) and a coherent coupling (energy $\hbar \omega_{\rm fs}$) between the two neutral exciton states. Decay processes are drawn with dashed lines. Bottom: The calculated nuclear spin dynamics as a function of laser detuning for a $\sigma^{+}$ pump, $+0.5\mbox{ T}$ external field, and the parameters described in the text. The solid (dashed) line shows the stable (unstable) solution for $\frac{d}{dt}\langle I_z \rangle=0$.}  
\label{figure4}                                       
\end{figure}

First, we calculate the effect of the laser field on the dynamics of a five-level system, consisting of the vacuum state $\ket{0}$, the two $X^{0}$ exciton states,  $\ket{\Uparrow \downarrow}$ and $\ket{\Downarrow \uparrow}$, and the two $X^{1-}$ states, $\ket{\Uparrow \downarrow \uparrow}$ and $\ket{\Downarrow \uparrow \downarrow}$. The laser is $\sigma^{+}$ polarized and drives the $\ket{0} \leftrightarrow \ket{\Uparrow \downarrow}$ but not the $\ket{0} \leftrightarrow \ket{\Downarrow \uparrow}$ transition on account of the selection rules. The optical Rabi energy is $\hbar \Omega$, the detuning $\hbar \delta = \hbar \omega -\hbar \omega_0$, where $\omega$ is the angular frequency of the laser and $\hbar \omega_0$ is the eigenenergy of $\ket{\Uparrow\downarrow}$ and $\ket{\Downarrow\uparrow}$ without magnetic field. Coupling between $\ket{\Uparrow \downarrow}$ and $\ket{\Downarrow \uparrow}$ is characterized by the fine structure $\hbar \omega_{\rm fs}$. Decay processes are sketched in the level diagram, Fig.\ \ref{figure4} (top). The neutral excitons can decay by spontaneous emission to $\ket{0}$ at rate $\tau_{0}^{-1}$; or they can become trion states via tunneling at rate $\tau_{\rm in}^{-1}$. Starting with the entire population in the ground state, we use the master equation for the density matrix to determine the occupation probabilities $p_{\ket{\Uparrow\downarrow\uparrow}}$ and $p_{\ket{\Downarrow\uparrow\downarrow}}$ of the trion states after time $\tau_{1}$, the spontaneous recombination lifetime of $X^{1-}$, resulting in the rates of creating an $\uparrow$, $\downarrow$ electron via optical recombination. 

After trion recombination, the free electron interacts with the $N$ quantum dot nuclei through the contact hyperfine interaction before it tunnels out at rate $\tau_{\rm out}^{-1}$. The spin flip-flop probability $p_{\rm ff}$ is
\begin{displaymath}
p_{\rm ff} = \frac{2\gamma\tau_{\rm out}^2}{\left(4\gamma + \xi\right)\tau_{\rm out}^2 + \hbar^2}, 
\end{displaymath}
where $\gamma = \frac{A^2}{4 N}\left(I - \left|\langle I_z \rangle\right|\right)$ and $\xi = \left(g_{\rm e} \mu_{\rm B} B_{z} + A\langle I_z \rangle \right)^2$, with $g_{\rm e}$ as the electron g-factor \cite{epaps}. The combination of electron creation rate and flip-flop probability results in a dynamic equation for the nuclear spin polarization. $\langle I_z \rangle$ is driven up depending on 
$p_{\ket{\Uparrow\downarrow\uparrow}}$, down depending on $p_{\ket{\Downarrow\uparrow\downarrow}}$, and decays in the absence of driving with rate $\Gamma_{\rm leak}$: 
\begin{displaymath}
\frac{d}{dt}\langle I_z \rangle \simeq \frac{p_{\rm ff}}{N \tau_1} \left[p_{\ket{\Uparrow\downarrow\uparrow}} - p_{\ket{\Downarrow\uparrow\downarrow}}\right]_{\mbox{\small $t=\tau_{1}$}} - \Gamma_{\rm leak} \langle I_z \rangle.
\end{displaymath}
We solve this equation numerically to find stable values of $\langle I_{z}\rangle$ as a function of laser detuning $\hbar\delta$. At each solution one can also calculate the Overhauser shift $\Delta_{\rm n}$ and the polarization degree $P$ in the quantum dot emission \cite{epaps}. 

Parameters are set by in situ characterization and by comparison with previous experiments, making small tweaks to fit the experimental data in Fig.\ \ref{figure2}. We use the following values \cite{epaps}: $\hbar \Omega=23\mbox{ $\mu$eV}$, $\hbar \omega_{\rm fs} = 40\mbox{ $\mu$eV}$, $\tau_{0}=0.75\mbox{ ns}$, $\tau_{1}=0.95\mbox{ ns}$, $\tau_{\rm in}=35\mbox{ ps}$, $\tau_{\rm out}=5\mbox{ ps}$, $N=8.5 \times 10^{4}$, $\Gamma_{\rm leak}=0.1\mbox{ s}^{-1}$ and $g_e=-0.5$. Fig.\ \ref{figure4} (bottom) contains a plot of $\frac{d}{dt}\langle I_z \rangle$, showing that the solution for $\langle I_z \rangle$ changes from positive to negative with a region of bistability. The calculated $P$ and $\Delta_{\rm n}$ are plotted in the right panels to Fig.\ \ref{figure2}. Close to the optical resonance, there is an excellent agreement with the experimental results. 

The theory offers an explanation for the counterintuitive inversion of the PL polarization. When the $\sigma^{+}$-polarized laser comes into resonance with the forbidden $\ket{0} \leftrightarrow \ket{\Downarrow \uparrow}$ transition, a combination of the allowed $\ket{0} \leftrightarrow \ket{\Uparrow \downarrow}$ transition and the $\ket{\Uparrow \downarrow} \leftrightarrow \ket{\Downarrow \uparrow}$ coupling causes the population to build up in the $\ket{\Downarrow \uparrow}$ state, leading to electron spin $\downarrow$ creation following tunneling in and recombination. When the laser is then tuned further, the allowed $\ket{0} \leftrightarrow \ket{\Uparrow \downarrow}$ transition takes over and the cycle results in the creation of electron spin $\uparrow$. The creation of a particular electron spin leads to nuclear spin polarization which alters the energies of the $\ket{\Downarrow \uparrow}$, $\ket{\Uparrow \downarrow}$ states via the Overhauser field. This feedback results in a bistability close to the forbidden transition and continuous tuning thereafter.

We have explored some of the parameter space theoretically. For parameters close to the ones used in this experiment, a region of bistability exists when $\Gamma_{\rm leak}$ is small enough. A bistability is definitely possible even at zero magnetic field, provided that $\Gamma_{\rm leak}\lesssim 1$ s$^{-1}$ and that the tunneling times are increased relative to those in this experiment. The inversion in polarization can be enhanced to at least $P=+0.85 \leftrightarrow -0.85$, again by increasing the tunneling times and also by optimizing the $\omega_{\rm fs}:\Omega$ ratio. Furthermore, at detunings larger than those at the bistability, these parameters allow continuous control of the $\ket{\Downarrow \uparrow}$, $\ket{\Uparrow \downarrow}$ eigenenergies from $\mp 40$ to $\pm 50$ $\mu$eV, and, following Refs.\ \cite{epaps,Vink}, a reduction in the variance of the nuclear spin distribution by factors $\sim 5$. All these features are attractive for spin qubits and single photon emitters.

We thank Bill Coish for very helpful discussions and acknowledge funding from Swiss NSF, DARPA QuEST, \mbox{EPSRC}, The Royal Society (BDG), and ANR QUAMOS (BU).

\end{document}